%% file: main.tex
\documentclass[%
 amsmath,amssymb,
 aps, physrev,
 twocolumn]{revtex4-2}
\usepackage{graphicx}
\usepackage{amsmath}
\usepackage{bigstrut}
\usepackage{hyperref}
\usepackage{placeins}
\usepackage{ulem}
\usepackage{color}

\usepackage{tikz}
\usetikzlibrary{graphs, positioning, arrows.meta, calc}

\newcommand*\plink{p_\mathrm{link}}
\newcommand*\prewire{p_\mathrm{rewire}}
\newcommand*\pnoise{p_\mathrm{noise}}
\newcommand*\meank{\langle k \rangle}

\begin{document}

\title{Collective decision-making with heterogeneous biases:\\Role of network topology and susceptibility}

\author{Yunus Sevinchan} 
    \affiliation{Institute for Theoretical Biology, Department of Biology, Humboldt Universit\"at zu Berlin, 10099 Berlin, Germany}
    \affiliation{Research Cluster of Excellence ‘Science of Intelligence’, 10587 Berlin, Germany}
\author{Petro Sarkanych} 
    \affiliation{Institute for Condensed Matter Physics of the National Academy of Sciences of Ukraine, Lviv, 79011, Ukraine}
    \affiliation{$\mathbb{L}^4$ Collaboration and Doctoral College for the Statistical Physics of Complex Systems, Lviv-Leipzig-Lorraine-Coventry, Europe}
\author{Abi Tenenbaum} 
    \affiliation{Yale University, Department of Physics, New Haven, Connecticut, USA}
\author{Yurij Holovatch} 
    \affiliation{Institute for Condensed Matter Physics of the National Academy of Sciences of Ukraine, Lviv, 79011, Ukraine}
    \affiliation{$\mathbb{L}^4$ Collaboration and Doctoral College for the Statistical Physics of Complex Systems, Lviv-Leipzig-Lorraine-Coventry, Europe}
    \affiliation{Centre for Fluid and Complex Systems, Coventry University, Coventry, CV1 5FB, UK}
	\affiliation{Complexity Science Hub Vienna, 1080 Vienna, Austria}
\author{Pawel Romanczuk}
    \affiliation{Institute for Theoretical Biology, Department of Biology, Humboldt Universit\"at zu Berlin, 10099 Berlin, Germany}
    \affiliation{Research Cluster of Excellence ‘Science of Intelligence’, 10587 Berlin, Germany~\vspace{1pc}}

\date{
    29 November 2024
}

\begin{abstract}
The ability of groups to make accurate collective decisions depends on a complex interplay of various factors, such as prior information, biases, social influence, and the structure of the interaction network.
Here, we investigate a spin model that accounts for heterogeneous preferences and enables control over the non-linearity of social interactions.
Building on previous results for complete graphs and regular 2D lattices, we investigate how the modification of network topology towards (sparse) random graphs can affect collective decision-making. We use two different measures of susceptibility to assess the responsiveness of the system to internal and external perturbations.
In particular, we investigate how the maximum of susceptibility depends on network connectivity.
Based on our findings, we discuss how collective systems might adapt to changes in environmental fluctuations by adjusting their network structure or the nature of their social interactions in order to remain in the region of maximal susceptibility.

\vspace{1pc}


\end{abstract}

\maketitle

\section{Introduction}
\label{sec:intro}

Collective decision-making is believed to be a core advantage of living in groups.
The pooling of information often enables collectives to make better choices than any individual could alone~\cite{krause2010swarm,sumpter2010collective}.
This ``wisdom of crowds'' effect can help group-living animals locate food~\cite{bhattacharya2014collective,mezey2024visual}, avoid predators~\cite{ward2008quorum,klamser2021collective}, or navigate challenging environments~\cite{guttal2010social}.
In human communities, collective decision-making is a core component of collective problem-solving, so understanding what conditions give rise to wisdom of crowds effects poses an important scientific problem~\cite{kao2014decision,navajas2018aggregated,tump2020wise,winklmayr2020wisdom}.
In general, this depends on a complex interplay of various factors including social influence and conformity, structure of the interaction network, prior information, and biases; despite decades of research many questions remain open.
Our work uses agent-based models of collective decision-making to generate deeper insights and uncover the specific mechanisms underlying wisdom of crowds effects.

Spin models provide a well-developed framework to study fundamental mechanisms in collective-decision making processes~\cite{redner2019reality,krapivsky2003dynamics}.
Recently, Hartnett~\textit{et al.} suggested a spin model that accounts for several important aspects of collective decision-making such as heterogeneous preferences (biases) and non-linear processes in social interactions~\cite{hartnett2016heterogeneous}.
They analysed the model on a 2D square lattice and showed that uninformed individuals with no biases can help the collective reach consensus.
In addition to the parameter modulating the strength of individual biases, the model incorporates a parameter that regulates the non-linearity of social influence, which significantly affects the dynamics of the system~\cite{hartnett2016heterogeneous}.
For limiting values of this non-linearity parameter, the (linear) voter model~\cite{redner2019reality} or the (dichotomous) majority rule model~\cite{krapivsky2003dynamics} is recovered.
This non-linearity parameter also implicitly controls the stochasticity of agents updating their opinion due to social influence.

Generally, collective decision-making processes in biological systems are also subject to quasi-random environmental influences that are independent of within-network social interactions. 
To account for this, we recently extended the original model proposed in~\cite{hartnett2016heterogeneous} by incorporating an independent noise process that acts on individual opinions.
We used the extended model to study how environmental noise affects consensus dynamics on a complete graph~\cite{Sarkanych2023physbiol,Sarkanych2024CMP}.

Another important aspect of the dynamics of a complex system is its structure, i.e. how the different components are connected with each other, which is captured by the interaction network.
Hartnett~\textit{et al.} introduced their model on a regular lattice~\cite{hartnett2016heterogeneous} and our recent work considered the model on a complete graph for analytical tractability~\cite{Sarkanych2023physbiol,Sarkanych2024CMP}.
However, typical real-world networks deviate from these limiting cases; instead of highly regular spatial structures (e.g. square 2D lattice), they tend to have additional long-range connections, and instead of all-to-all connected systems (complete graphs), they tend to have a limited number of connections per agent.

In the present work, we investigate how changes in network topology (starting from a complete graph~\cite{Sarkanych2024CMP} or a 2D~lattice~\cite{hartnett2016heterogeneous}) affect the behaviour of the extended Hartnett model~\cite{Sarkanych2024CMP}. 
Specifically, we study two different mechanisms for introducing deviations in network topology: (i)~randomly removing links from a fully connected graph, and (ii)~randomly rewiring links from nearest neighbours on a 2D~lattice to random agents within the system~\cite{watts1998collective}.
The parameters controlling the graph generation are the linking probability~$\plink$ and the rewiring probability~$\prewire$.
Applying either mechanism with $\plink < 1$ or $\prewire \to 1$, respectively, yields an Erd\"{o}s-R\'enyi~(ER) random graph.

Furthermore, we quantify susceptibility (using two different measures) in order to study specifically how network topology affects the system's sensitivity to external influence.
Motivated by experimental observations, it has been suggested that real-world systems should operate near critical points, where susceptibility becomes maximal~\cite{mora2011biological,GomezNavaNatPhysFish}.
It is therefore interesting to understand how changes in the structure of the interaction network affect the location of critical points, as this may be a primary mechanism by which real-world systems can modify their distance to criticality and thus increase their environmental sensitivity~\cite{romanczuk2023phase}.

The rest of the paper is structured as follows: In Section \ref{sec:model_description} we describe the model and the parameters and also define the observables that will be used to characterise the system.
In Section~\ref{sec:model_impl} we describe the implementation of the model for performing computer simulations.
We present the main results for the random graph with varying linking probability in Section~\ref{sec:results_ER}; in Section~\ref{sec:results_WSG} we focus on the rewiring process in a 2D~lattice.
Finally, in Section~\ref{sec:adaptation}, we discuss how a system might adapt to changing environmental conditions in order to retain maximum susceptibility and in Section~\ref{sec:conclusions}, we offer our concluding remarks.

\section{Model Description}\label{sec:model_description}

Here, we briefly describe the model from Ref.~\cite{hartnett2016heterogeneous} with the additional external stochastic noise process~\cite{Sarkanych2023physbiol}.

Each agent in a group of $N$ individuals is described by a spin variable $S_i=\pm 1$, $i=1,\dots,N$.
The internal biases of an individual are regulated by a variable $\omega_i$, where $\omega_i>1$ corresponds to the individuals that are biased towards the state $S_i=+1$, while $0 < \omega_i < 1$ describe the individuals biased toward $S_i=-1$. 
Agents with $\omega_i=1$ have no bias and are called unbiased individuals.
The distribution of biases is assumed to be random and uniform, with $\rho_0$ denoting the fraction of unbiased individuals, and $\rho_+$ and $\rho_-$ denoting the fraction among biased agents toward $S_i=+1$ and $S_i=-1$, respectively, such that $\rho_+ + \rho_- = 1$.
This leads to the natural normalisation condition
\begin{equation}
    \label{eq:rho_normalisation}
    \rho_0+(1 - \rho_0)\rho_+ + (1 - \rho_0)\rho_-=1 \, .
\end{equation}
Each of the individuals is affected by the so-called social field exerted by its nearest neighbours,
\begin{equation}
    \label{eq:social_field}
    h_i=\frac{\omega_i n_i^+ - n_i^-}{\omega_i n_i^+ + n_i^-}\, ,
\end{equation}
where $n_i^\pm$ is the number of nearest neighbours of agent~$i$ with opinions $\pm 1$.
As one can see from Eq.~\ref{eq:social_field}, the social field depends on the individual bias -- it is strengthened if the agent's bias coincides with the dominating opinion among its neighbours and weakened otherwise.

At each step of model iteration, the next state of an agent is defined by a sigmoid function: 
\begin{equation}
    \label{eq:flipping_prob}
    G_i=\frac{1}{2} \left( 1 + \frac{\tanh (b h_i)}{\tanh (b)}\right)\, ,
\end{equation}
where $b$~is the so-called non-linearity parameter.
If $S_i = -1$, the next state will be $+1$ with probability $G_i$; for $S_i = +1$, it will flip to $-1$ with probability $1-G_i$.
Additionally, we add an independent and external stochastic noise process.
At each iteration step, there is a probability $\pnoise$ that an agent will switch to the opposite state. 
This stochastic noise has a similar effect as temperature has in statistical physics models~\cite{Sarkanych2023physbiol}.

\medskip

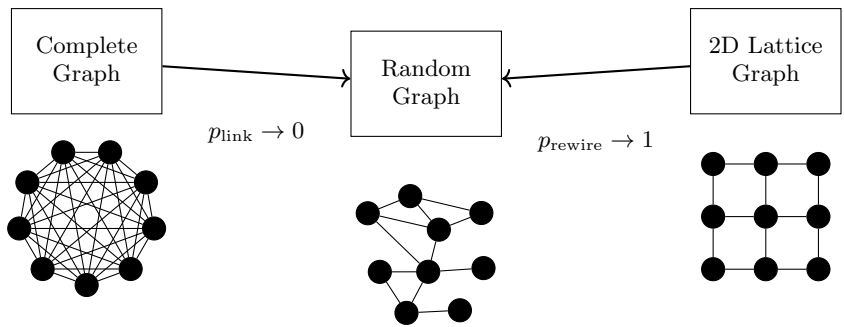
\begin{figure*}[t!]
    \centering
    \input{figures/schematic.tikz}
    \caption{
        \label{fig:schematic}
        Schematic of the two different graph generation mechanisms we study:
        By reducing the linking probability~$\plink$, a complete graph is pruned into a random graph of decreasing link density;
        by rewiring links from nearest neighbours to random nodes with probability~$\prewire$, a regular 2D~lattice graph is transformed into a random graph with a corresponding mean degree.
        Note that, for simplicity, the shown lattice graph has no links on the boundary -- however, the graphs we studied include links from high- to low-value boundaries, such that all nodes have the same mean degree, alleviating boundary effects.
    }
\end{figure*}

As mentioned in the Introduction and illustrated in Fig.~\ref{fig:schematic}, we investigate two different mechanisms to generate the interaction networks:
\begin{itemize}
    \item Changing the linking probability~$\plink$:
    In the limiting case $\plink=1$, one obtains the complete graph, previously analysed in Ref.~\cite{Sarkanych2023physbiol}, while for $\plink<1$ the resulting network is an Erd\"{o}s-R\'enyi random graph with mean node degree depending on $\plink$.
    \item Rewiring of regular 2D~lattice links with probability~$\prewire$:
    For $\prewire=1$, one obtains a random graph analogous to the graph from the previous case with the corresponding mean degree.
    We will consider two initial configurations of a 2D lattice: square lattice (4-neighbourhood) and square lattice with diagonal interactions (8-neighbourhood), with mean degrees~$\meank \in \{4, 8\}$, respectively.
\end{itemize}
The lattice topology is similar to the one considered in Ref.~\cite{hartnett2016heterogeneous}, however, our results are not directly comparable for two reasons:
Firstly, we consider a model with external noise that affects the opinion formation process.
Secondly, when sweeping through the model parameters to analyse how they affect the consensus, we set the bias-related parameters ($\omega_\pm$) to constant values, while in Ref.~\cite{hartnett2016heterogeneous} they are chosen such that the consensus opinion is constant.

\medskip

For a group of animals to be adaptive, they must be highly responsive to changes in external conditions. 
This may be reflected by the observation that some biological systems appear to operate near the critical point of a phase transition between two qualitatively different behavioural regimes~\cite{mora2011biological,GomezNavaNatPhysFish}.
For the considered decision-making model, such a critical point would be analogous to high values of susceptibility in a spin model.
When doing simulations of equilibrium statistical models, there are two established ways to quantify susceptibility:
The first is through measurement of the variance of magnetisations, which in our case is the variance of the equilibrium opinion~$m$:
\begin{equation}
    \label{eq:var}
    \mathrm{var}(m)=\langle m^2\rangle-\langle m\rangle^2 \quad, 
\end{equation}
with $m=\frac{1}{N}\sum_{i=1}^N S_i$.

An alternative way to define the magnetic susceptibility is by introducing a negligibly small external field $B$. 
In our model, this field is realized as an additional probability for a spin to flip to a specific state.
In theory, this definition uses the derivative $\chi=\partial m / \partial B$.
However, in simulations we have to set small but finite values of the external field; in practice, this will also be a constant value.
Thus, we are using the following physics-inspired definition
\begin{equation}
    \label{eq:chi}
    \chi = m(B)-m(B=0) \quad.
\end{equation}

For systems where the fluctuation-dissipation theorem~\cite{kubo1966fluctuation} holds, these two definitions should provide analogous outcomes.
However, the model under consideration is not in the equilibrium state and has a finite size.
Both of these factors may result in deviations between these two susceptibility measures.

\section{Model implementation}\label{sec:model_impl}

Numerical simulations were performed using an extended implementation of the one presented in Ref.~\cite{Sarkanych2024CMP}.
The model was implemented in modern~C++ and uses a graph data structure based on the Boost Graph Library~\cite{BGL} to represent arbitrary graphs.
We used the \textit{Utopia} modelling framework~\cite{Riedel20,Sevinchan20,Sevinchan20boosting} for the implementation of the model, simulation configuration, parallelised parameter sweeps, as well as efficient reading, writing, and evaluation of high-dimensional simulation output.
All code is available online~\cite{Sevinchan24model}.

At the beginning of each model simulation, an undirected graph~$g$ is generated according to the specified graph parameters.
All studied graphs had a size of $N=10^4$ agents, but their linking mechanisms and hence topology varied depending on the selected graph generator (described below) and the chosen parameters.
Agents are then grouped according to $\rho_0$ and $\rho_\pm$ and respective bias values $\omega_0 = 1$ or $\omega_\pm$ are assigned.
Each agent is randomly assigned an opinion~$S_i$ such that the system's mean magnetisation matches a specified $m_0 := m(t=0)$ initial magnetisation value.
The initial magnetisation~$m_0$ for the different parameter scans performed was chosen such that relaxation towards meta-stable states, as described in~\cite{Sarkanych2023physbiol}, is reduced whenever possible.
For the selected bias scenarios this amounts to $m_0 = +1$.

During model iteration, each agents' social field~$h_i$ is computed using Eq.~(\ref{eq:social_field}).
Depending on their current state, agents will flip to $S_i=+1$ with a probability of~$G_i$ (see Eq.~(\ref{eq:flipping_prob})) or to $S_i=-1$ with a probability of~$1-G_i$.
Subsequently, through an additional and independent noise process, agents have a probability of $\pnoise \in [0, 1]$, also referred to as noise level, to flip to the opposite state, regardless of whether they previously flipped due to the social field.
Finally, all state changes are applied synchronously and the model continues to the next iteration step.

Iteration is repeated until the specified number of iteration steps is reached, concluding one realisation.
For the next realisation, the agents' biases and states are randomly re-initialised on the same graph and the above procedure is repeated a total of $M_\text{rep}$ times.
To ensure that variations in the generated graph structure are also taken into account, we additionally regenerate graphs using $M_\text{seeds}$ different random number generator seeds.

For the line plots, $M_\text{rep} = 32$~repetitions were performed for each parameter combination and we used $M_\text{seeds} = 16$ different seeds, resulting in a total number of $M_\text{tot} = M_\text{rep} \cdot M_\text{seeds} = 512$ realisations per parameter combination.
For the heatmap plots, we reduced the number of realisations per parameter combination to counteract the higher number of values along the $\plink$ or $\prewire$ parameter dimension and retain an appropriate simulation time:
We chose $M_\text{tot} = 96$, with ($M_\text{rep} = 24$, $M_\text{seeds} = 4$) for the ER~graph and ($M_\text{rep} = 16$, $M_\text{seeds} = 6$) for the rewired lattice graph.

\medskip

To numerically determine the final state magnetisation, we ran simulations sufficiently long and used values from the end of the resulting time series~$m(t)$ to compute the final state magnetisation.
With the ER topology, systems typically reached their pseudo-steady state within at most 200~iteration steps, so we chose a simulation time of 1000~steps for these graphs.
Due to the more localised interactions in the lattice graph of size $100^2$, reaching the pseudo-steady state took longer in these systems, which is why a simulation time of 5000~steps was chosen.
To reduce the effect of the external noise process on quantitative results, we averaged the magnetisation~$m$ over the last 50~steps to arrive at the mean final state magnetisation.
The final state variance, see Eq.~(\ref{eq:var}), is computed over the same last 50~simulation steps of each realisation.

For the determination of the magnetic susceptibility~$\chi$ according to Eq.~(\ref{eq:chi}), we run simulations again but apply a small external field in the following way: after the noise process, we add an additional stochastic process where an agent flips to $S_i = +1$ with a probability~$B$ regardless of its current state.
The asymmetry of this process represents an external field aligned with the $+1$ spin direction where $B$ corresponds to the field strength.
We choose $B=0.01$ and note that the results are qualitatively similar for smaller values, but fluctuations of~$\chi$ increase and thus require more realisations to yield a clear signal.

\medskip

Erd\"{o}s-R\'enyi (ER) random graphs $g = g(N,\plink)$ were generated by adding each possible undirected edge $(u, v) = (v, u)$ between two agents $u$ and $v$ with a probability of $\plink \in [0, 1]$.
A linking probability $\plink = 1$ thus yields a complete graph, in which case a mean-field approach can be used that optimises the simulation by avoiding the costly iteration over all $\mathcal{O}(N^2)$ edges; see Ref.~\cite{Sarkanych2024CMP} for more details.

The second graph topology we investigated was based on 2D~regular lattice graphs of size $L^2 = N$ and with a degree of $k \in \{4, 8\}$.
To avoid boundary effects, a toroidal embedding was chosen, such that all nodes have an identical neighbourhood.
Subsequently, the regular lattice structure was changed by randomly rewiring edges:
With a probability of $\prewire$, every undirected edge $(u, v) = (v, u)$ is rewired to $(u, w) = (w, u)$.
The node~$w$ was chosen randomly among those not already connected to node~$u$; furthermore, self-edges $(u, u)$ were excluded.
This graph topology is hence described by $g = g(N,k,\prewire)$.

With increasing $\prewire$, the purely local interactions on the lattice are incrementally replaced with long-range interactions, effectively reducing the diameter of the graph.
No edges were added or removed, meaning that $\prewire = 1$ yields a random graph with mean degree $\meank \in \{4, 8\}$, which is identical to an ER~random graph with a linking probability~$\plink = 2 \meank / N$.
For the chosen lattice neighbourhood sizes of 4 and 8, the corresponding linking probabilities are $\plink \in \{ 0.0008, 0.0016 \}$, respectively.
This procedure is equivalent to Watts-Strogratz~(WS) graph generation~\cite{watts1998collective}, but instead of using a regular ring graph as a starting point for rewiring, we used a toroidal lattice graph.

\section{Impact of linking probability}
\label{sec:results_ER}

We start our analysis from an ER random graph where two arbitrary nodes are connected with a probability $\plink$.
In the limiting case of $\plink=1$, one arrives at the same results as for the complete graph~\cite{Sarkanych2024CMP}.
It was shown that even for the case of the complete graph, the phase diagram is very rich and the critical behaviour is strongly dependent on model parameters like the bias configuration.
For the purpose of this paper, we will be focusing on a single bias configuration described in Ref.~\cite{Sarkanych2024CMP}, consisting of two groups with weak and slightly asymmetric biases: $\omega_+=1.5, \omega_-=0.7, \rho_+=0.6$ and $\rho_-=1-\rho_+=0.4$.
If not mentioned otherwise, we used a fixed value of $b=1.0$ for the non-linearity parameter in Eq.~(\ref{eq:flipping_prob}), corresponding to an almost linear mapping between $h_i$ and~$G_i$.

\begin{figure}[t!]
    \centering
    \includegraphics[width=1\linewidth]{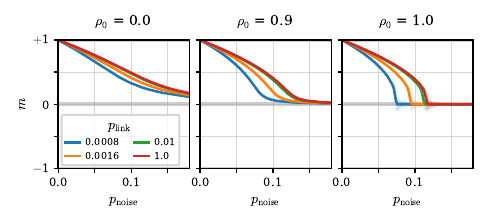}
    \caption{\label{fig_er_magnetization}
        Equilibrium opinion~$m$ for three different linking probabilities $\plink$ (colours) of an ER~random graph and three different bias group sizes~$\rho_0$ (columns).
        Shaded areas denote the standard deviation across 512~realisations for each parameter combination.
        $\plink = 1$ represents a complete graph.
        As one can see, as linking probability increases, the critical point for the completely unbiased case, $\rho_0 = 1$, moves to higher noise levels.
        For biased cases, $\rho_0 < 1$, higher linking probability has a similar effect, but the critical point is not observed.
    }
\end{figure}

In Fig.~\ref{fig_er_magnetization} we present the equilibrium opinion~$m$ as a function of the external stochastic noise for different linking probabilities and three different densities of biased individuals.
Only in the case, where all individuals are unbiased, one observes a critical point; this point is the same as for the ordinary Ising model on a random graph.
When biased individuals are added, $\rho_0 < 1$, the behaviour is closer to that of a system in an external magnetic field.
In addition, when biased individuals are added, the curve moves upward, similarly to how a magnetic system would react to an increase in magnetic field.
As linking probability increases, the values of the critical noise level increases as well, which corresponds to
the increase of critical temperature on the random graph \cite{dorogovtsev2008critical,krasnytska2013}.
However, since the noise process, controlled by $\pnoise$, is not fully equivalent to temperature, the dependency is not linear.
Notably, the critical point of a graph with $\plink = 0.01$ is already very close to that of the complete graph; most of the change in the critical noise level occurs for $0 < \plink < 0.01$.

\begin{figure}[t!]
    \centering
    \includegraphics[width=\linewidth]{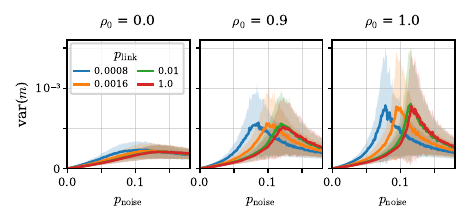}
    \caption{\label{fig_er_variance}
        The variance of the equilibrium opinion~$m$ as a measure of susceptibility, corresponding to Fig.~\ref{fig_er_magnetization}.
        For systems with many unbiased individuals, the peak of the response function is clearly visible, while for the system where most of the agents are biased, peaks are less pronounced.
        As the connectivity increases, the position of the maximum variance peak shifts to higher noise levels.
    }
\end{figure}

Next, we analyse the variance of the equilibrium opinion as one of the approaches to study the system's susceptibility.
In Fig.~\ref{fig_er_variance} the variance of the consensus opinion, as defined by Eq.~(\ref{eq:var}), is shown as a function of external stochastic noise for different connectivities and biases.
It is evident that the peak of the variance shifts to higher noise levels as the linking probability increases, coinciding with the critical noise level in the consensus opinion, see Fig.~\ref{fig_er_magnetization}.
This behaviour is similar to critical phenomena on random networks where critical temperature is proportional to the linking probability \cite{dorogovtsev2008critical,krasnytska2013}.
On the other hand, there is not much of a difference between the heights of the peaks for different linking probabilities.
While changes in linking probability do not seem to affect the height of the variance peaks, peaks are higher with a higher fraction of unbiased individuals~$\rho_0$. We note that, although a (quasi-)critical point exists only in the fully unbiased case of $\rho_0=1$, we observe well-defined maxima of susceptibility also for (fully) biased systems with $\rho_0=0$.  
The susceptibility curves are asymmetric with respect to the maximum. The left side of the curves is steeper than the right; this asymmetry seems not to be affected by changes in connectivity or the percentage of biased individuals.

\begin{figure}[t!]
    \centering
    \includegraphics[width=1\linewidth]{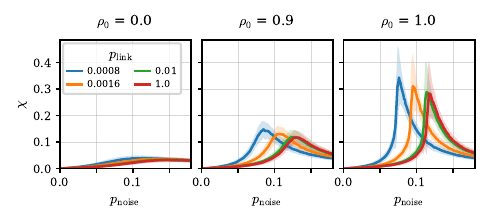}
    \caption{\label{fig_er_chi}
        Change in equilibrium opinion in response to a small external magnetic field, corresponding to Fig.~\ref{fig_er_magnetization}.
        A clear peak of the susceptibility is only observed for the systems with a high~$\rho_0$.
        As the connectivity increases, the maximum of the susceptibility shifts to higher noise levels.
    }
\end{figure}

In Fig.~\ref{fig_er_chi} we show the susceptibility measured as the response $\chi$ of the system to a small external magnetic field~$B$, see Eq.~(\ref{eq:chi}).
As in the plot of the stable state variance, peaks are more pronounced for a higher number of unbiased individuals and peak positions change to higher noise levels for higher linking probabilities; again, for the unbiased system, the peak coincides with the position of the critical point.
Alongside the shift in the position of the peaks, their height slightly decreases with increasing linking probability, which is unlike the observations in the variance plot, where peaks are of equal height.
This suggests a slight reduction in responsiveness of the better connected networks to the external field.

\section{Impact of rewiring}
\label{sec:results_WSG}

In this section, we consider our model on a 2D~lattice with periodic boundary conditions and a certain portion of the links~$\prewire$ randomly rewired.
As the fraction of rewired links reaches $\prewire=1$ one recovers the random graph with the same mean degree~$\meank$ as the number of interacting neighbours on the lattice.
For instance, with the nearest neighbours square lattice as a starting point, the mean degree is $\meank=4$.
Notably, even a small portion of rewired links significantly affects the topology of the system since rewiring adds, on average, long-range interactions to the system, causing the small-world effect \cite{watts1998collective}.
Below, we consider the same bias configuration as in the previous section and study two lattice configurations where agents are interacting with~4 or~8 neighbours.

\begin{figure}[t!]
    \centering
    \includegraphics[width=\linewidth]{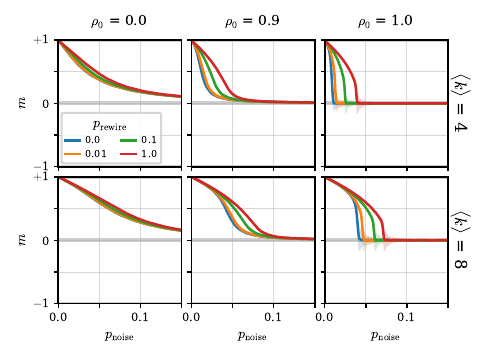}
    \caption{\label{fig_wsg_magnetization}
        Equilibrium opinion as a function of external noise levels for a rewired 2D~lattice graph with different values for $\prewire \in \{ 0, 0.01, 0.1, 1 \}$ (line colours), fraction of unbiased individuals~$\rho_0$ (columns), and mean number of interacting neighbors $\meank$ (rows).
        The blue line represents a regular 2D~lattice graph, while the red line represents a fully rewired graph that corresponds to an ER~random graph with $\plink \in \{0.0008, 0.0016 \}$, respectively.
        Shaded areas denote the standard deviation across the 512~realisations performed for each parameter combination.
        Similarly to the case of the ER~random graph, the introduction of biases removes a clear second order phase transition.
        Also, as $\prewire$ or $\meank$ increases, the critical noise level moves to higher values.
    }
\end{figure}

In Fig.~\ref{fig_wsg_magnetization} we show the equilibrium opinion~$m$ depending on the external stochastic noise level for different rewiring probabilities and interaction radii.
In the completely biased case, $\rho_0=0$, the plot looks similar to that of the Ising model in an external magnetic field.
In the fully unbiased case, $\rho_0=1$, a second order phase transition can be observed.
As the number of interacting neighbours increases, the values for the critical noise levels increase as well; this behaviour is well known for mean-field-like models~\cite{krasnytska2013,dorogovtsev2008critical}.
By rewiring links, a similar effect is achieved, with critical noise levels increasing; this is due to the fact that rewired links introduce long-range correlations to the system, making it harder to break emerging ordering.
Changes in critical noise levels appear only for $\prewire \gtrapprox 10^{-2}$; for small $\prewire$, lattice-like interactions still dominate over the few random long-range links introduced through rewiring, thus having only little effect on the system as a whole.

\begin{figure}[t!]
    \centering
    \includegraphics[width=1\linewidth]{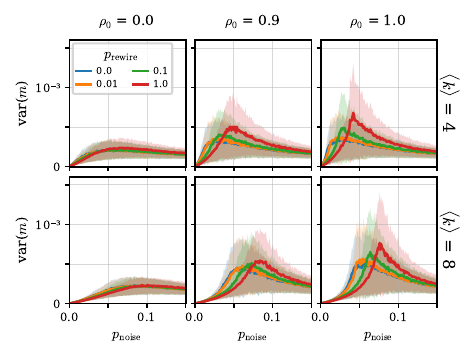}
    \caption{\label{fig_wsg_variance}
        Variance of the equilibrium opinion as a function of noise level, corresponding to Fig.~\ref{fig_wsg_magnetization}.
        An increase in the number of interacting neighbours or rewiring probability both make the peaks of the variance more pronounced and shift them to higher noise values.   
        Unlike in the corresponding observations when changing $\plink$, see Fig.~\ref{fig_er_variance}, the peaks increase in height when increaasing $\prewire$.
    }
\end{figure}

Next, we again examine the response functions by looking at the variance of the equilibrium opinion; in Fig.~\ref{fig_wsg_variance} this first susceptibility measure is shown.
As the number of interacting neighbours~$\meank$ increases, the variance peaks become more pronounced (in particular at low $\prewire$) and move to higher noise levels, which for $\rho_0=1$ corresponds to a change in the critical noise level.
Increasing the rewiring probability has a similar effect, but additionally leads to higher peak values -- this is unlike in the case of the ER~random graph, where changes in $\plink$ did not change the height of the variance peaks.
We speculate this may be due to the decreasing graph diameter when adding long-range interactions: on a lattice-like graph with $\prewire \approx 0$ the graph diameter is large and fluctuations first need to propagate through the lattice and may cancel each other out, while fluctuations on small-diameter random graphs, $\prewire \approx 1$, travel fast, thus causing more flips and a higher variance.

\begin{figure}[t!]
    \centering
    \includegraphics[width=1\linewidth]{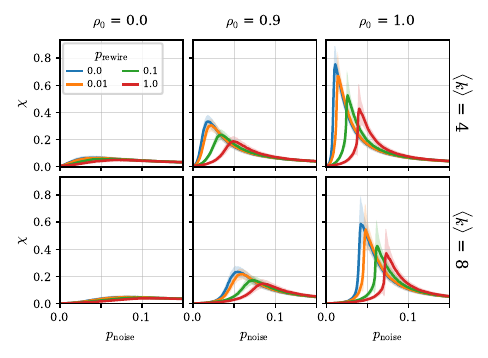}
    \caption{\label{fig_wsg_chi}
        Response of the equilibrium opinion to a small external field, corresponding to Fig.~\ref{fig_wsg_magnetization}.
        With increased rewiring or~$\meank$, peaks move to higher noise levels, but also decrease in height.
    }
\end{figure}

Finally, we consider the response of the equilibrium opinion to a small external magnetic field~$B$, equivalent to how it was done in the previous section for the ER~random graph.
In Fig.~\ref{fig_wsg_chi} we show the response~$\chi$ as a function of external noise level~$\pnoise$ for different rewiring probabilities and interaction radii.
The behaviour here is very similar to that of the variance in Fig.~\ref{fig_wsg_variance} with the sole exception that the height of the peak decreases as the rewiring probability increases.
This is directly opposite to the beheavior of the variance peaks, but consistent with the dependence of peak susceptibility on external fields observed for the ER~random graph, see Fig.~\ref{fig_er_chi}.

\section{Mechanisms to maintain maximum susceptibility}
\label{sec:adaptation}

In the model we studied here, the external stochastic noise can be interpreted as the influence of the environment on the system.
In principle, the intensity of this effect may vary over time or in reaction to the state of the system itself.
Assuming that there are benefits for systems to be in a state where they are highly sensitive to external input~\cite{munoz2018}, this raises the question of how such a system would need to \textit{adapt} to changes in the environment in order to \textit{retain} such a state of maximum susceptibility.
The observations from our model simulations indicate how changes in topology affect the position of these high-susceptibility manifolds.
In the previous sections we showed that there can be peaks in susceptibility despite the system not exhibiting a clear phase transition and quasi-critical point.
In this section, we explore how possible adaptations in topology or interaction mechanisms can counteract changes in environmental noise and keep the system in a state of high susceptibility.
Note that, to this end, we are not modelling the dynamic transition and temporal dynamics between one consensus state and another, but are quantifying how changes of the involved parameters are affecting the response functions of the system, and in particular how the change in $\pnoise$ as a core control parameter can be counterbalanced by changes in interaction parameters (network topology and non-linearity of social interactions).

\medskip

\begin{figure}[t!]
    \centering
    \includegraphics[width=1\linewidth]{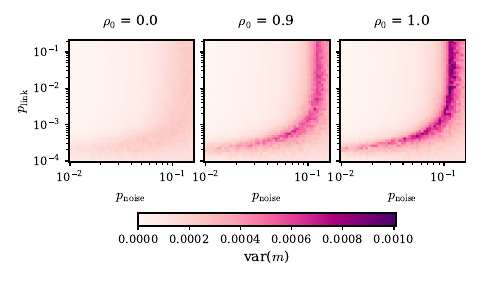}
    \includegraphics[width=1\linewidth]{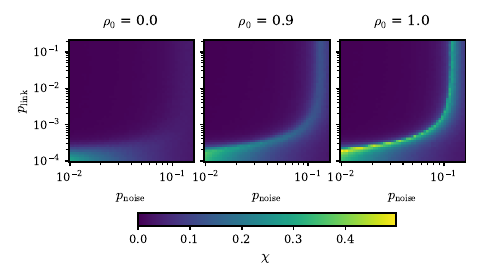}
    \caption{
        \label{heatmap_er}
        Heatmaps of the two response functions for the model on an ER random graph. The locations of the peaks are the same in both cases.
        As the linking probability increases, the noise level that corresponds to the peak increases as well until it reaches a ``plateau" at $\plink \gtrapprox 10^{-2}$.
    }
\end{figure}

In Fig.~\ref{heatmap_er} the response functions for different noise levels and linking probabilities are shown as heatmaps. 
It is evident that as the linking probability increases the noise level corresponding to the maximum value of susceptibility also increases.
This is true only for small values of the linking probability.
For increasing noise levels, the system would remain at maximum susceptibility if it would increase its link density by adding random links, i.e. increase~$\plink$.
The simple explanation is that the system is tightly connected for higher linking probabilities, making it harder to break the ordering.
However, this is not the case for very high noise levels, where increasing~$\plink$ alone does not affect the response functions any more.
In real-world systems, such an adaptation can be realised locally for each interacting individual by modifying the number of neighbouring opinions they take into account.

\medskip

\begin{figure}[t!]
    \centering
    \includegraphics[width=1\linewidth]{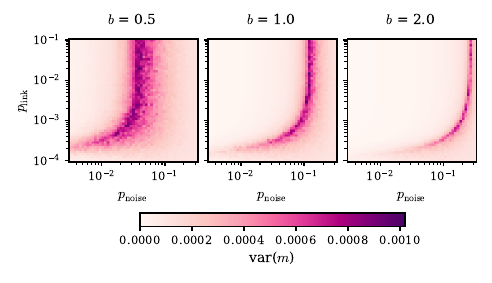}
    \includegraphics[width=1\linewidth]{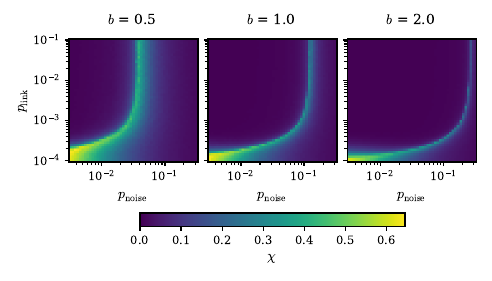}
    \caption{\label{heatmap_er_beta}
        Heatmaps of the two response functions for the model on an ER random graph and different non-linearity parameters~$b$.
        An increase in~$b$ also shifts the peak of response functions to higher noise levels.
        Note the log-scaled x-axis, letting the width of the peak appear wider when at lower values of~$\pnoise$.
    }
\end{figure}

Beside connectivity, the nature of the individual decision-making processes can also affect macroscopic susceptibility.
One important process is the reaction to an agents' social field, Eq.~(\ref{eq:flipping_prob}), which is defined by a sigmoidal function with the non-linearity parameter~$b$.
Given the shape of this function, $b$~can be interpreted as compliance with the neighbours' opinions, or conformity to the locally stronger opinion~\cite{Sarkanych2024CMP}:
The higher~$b$, the higher the sensitivity of individuals to perceived differences in local opinion.
In other words: with higher~$b$, a weak local majority becomes more likely to change opinion toward this local majority opinion.

In Fig.~\ref{heatmap_er_beta} we show how the non-linearity parameter~$b$ affects the location of the susceptibility peaks.
When~$b$ is increased, the peaks move to higher noise levels, the effect being similar to that of an increase in linking probability.
Thus, one adaptation that would keep the system at maximum susceptibility under increasing noise would be to make a change in the interaction mechanism: increasing the value of the non-linearity parameter~$b$.

\medskip

These two adaptation mechanisms rely on a local change in the interacting agent.
In the first case, it means increasing connectivity with other agents.
In the second case, it means to be more compliant with the neighbourhood.
Both these adaptations incur tighter connection with the rest of the system: topological for the first mechanism and dynamical for the second.

\medskip

\begin{figure}[t!]
    \centering
    \includegraphics[width=1\linewidth]{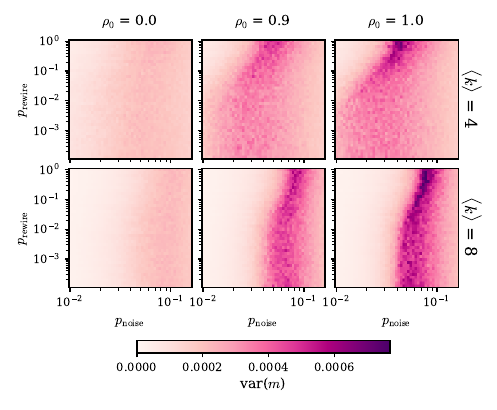}
    \includegraphics[width=1\linewidth]{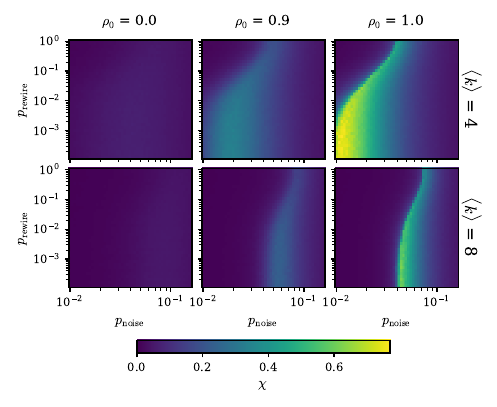}
    \caption{\label{heatmap_wsg_var_and_chi}
        Heatmaps of the equilibrium opinion variance (top) and response to a small external field (bottom) for the model on a 2D~rewired lattice.
        The strongest change in the peaks of the response functions is observed for rewiring probabilities larger than~$\prewire \approx 10^{-2}$.
        Note the double-logarithmic scale and that the range of simulated $\prewire$ values starts at $\prewire \approx 10^{-4}$, which is not exactly zero but with $N=10^4$ agents constitutes a negligible number of rewired links.
    }
\end{figure}

Next, we take a look at the rewired 2D~lattice again.
In Fig.~\ref{heatmap_wsg_var_and_chi} the consensus opinion variance and response to an external field is shown in a heatmap.
As before, a maximum susceptibility line as a function of external noise and rewiring probability can be seen.

When all the individuals are unbiased, there is a clear maximum line, but adding even a small portion of biased individuals smears it out; something that was already visible in Figs.~\ref{fig_wsg_variance} and~\ref{fig_wsg_chi}.
For both susceptibility measures, the response function is steeper for noise values below the peak than above it.
This is similar to the susceptibility of a second order phase transition that forms so-called $\lambda-$shape.
When the mean degree increases the peaks move to the higher noise values and the effect of rewiring becomes less pronounced.

Applying the same considerations regarding counteracting an increase in environmental noise to retain maximum susceptibility, such a lattice-based system could increase its rewiring, adding long-range interactions, and hence having its susceptibility peak at higher noise levels.

\medskip

\begin{figure}[t!]
    \centering
    \includegraphics[width=1\linewidth]{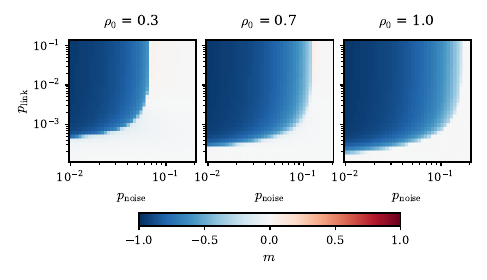}
    \includegraphics[width=1\linewidth]{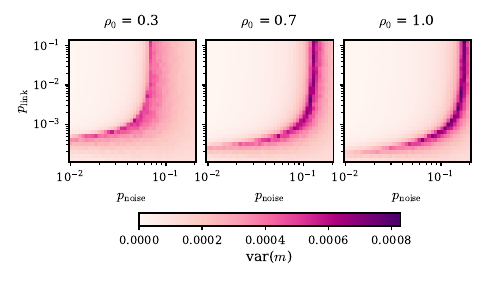}
    \includegraphics[width=1\linewidth]{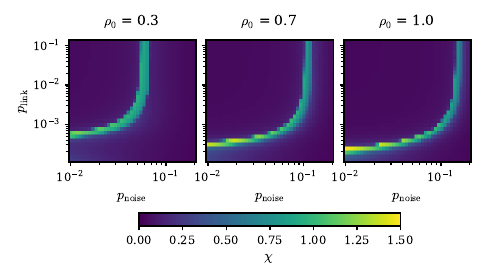}
    \caption{\label{heatmap_first_order}
        Heatmaps of the consensus opinion and response functions for a bias configuration where first order phase transitions are observed ($\rho_+=0.6$, $\rho_-=0.4$, $\omega_+=2$, $\omega_-=0.3$; mean values of 192~realisations for each parameter combination).
        The effects observed for the weakly biased configuration are present here as well; in addition, the position of the critical manifold is affected by the fraction of unbiased individuals~$\rho_0$, shifting towards smaller noise levels for lower~$\rho_0$.
    }
\end{figure}

Finally, after having discussed three possible mechanisms of adaptation, we shift our attention to other possible critical behaviours discussed in Ref.~\cite{Sarkanych2024CMP}.
In particular, we consider the case with a weakly-biased majority group and a strongly-biased minority group ($\rho_+=0.6$, $\rho_-=0.4$, $\omega_+=2$ and $\omega_-=0.3$) where first order phase transitions are observed for a certain range of $\rho_0$ values, not only for the completely unbiased case.

In Fig. \ref{heatmap_first_order} we show the heatmaps for the consensus opinion and the susceptibility measures for this case.
As one can see from the plots, the above-mentioned mechanisms also have a similar effect here. 
The only difference is that with the first-order transitions, peaks in the response functions are much more pronounced.
These figures also show that the choice of biases moves the line of maximum susceptibility, with more biased individuals (i.e. lower $\rho_0$) moving it to lower noise levels.
In other words: to maintain a consensus opinion near maximum susceptibility under increasing noise, the system would need to become less biased.
These adaptations could also occur on the level of individual agents.
We studied different bias configurations in more detail already in previous work, see Ref.~\cite{Sarkanych2024CMP}.

\section{Discussion \& Conclusions}
\label{sec:conclusions}

In this paper, we analysed the Hartnett model~\cite{hartnett2016heterogeneous} with an additional stochastic noise process acting on individual opinions~\cite{Sarkanych2023physbiol,Sarkanych2024CMP}. 
In addition to heterogeneity in the form of individual bias, the extended model lets us study the effect that environmental noise could have on the decision-making process.
We considered this model on Erd\"{o}s-R\'enyi random graphs, which play an important role in real-world interaction networks in addition to the previously studied complete graphs and spatial 2D~lattices~\cite{hartnett2016heterogeneous,Sarkanych2024CMP,Sarkanych2023physbiol}.
In order to link our analysis with past work, we investigated two different graph-generating mechanisms, continuously transforming the structures into one another:
Starting from the complete graph or the 2D~lattice, we respectively pruned or rewired links to change the topology of the interaction network.
This continuous approach let us systematically analyse how changes in topology and non-linearity of social interactions affected the phase diagram of the system generally and in particular the regions of maximal susceptibility, including quasi-critical points.

Recent experimental observations suggest that distributed biological systems, including animal groups, may operate near critical points \cite{mora2011biological,attanasi2014finite,gelblum2015ant,feinerman2018physics,poel2022subcritical,GomezNavaNatPhysFish}, a region of operation which has been linked to high responsiveness and the ability to make accurate decisions~\cite{mora2011biological,munoz2018,klamser2021collective,poel2022subcritical}. 
However, changes in environmental conditions might drive the system away from the critical region, diminishing this advantage. 
Therefore, to remain in a high susceptibility region (near the critical point), the system must adapt its parameters.

\medskip

To study these mechanisms, we used two definitions of susceptibility: the variance of the order parameter and the change in the order parameter as a response to an external field. 
For equilibrium systems, these two measures are equivalent according to the fluctuation-dissipation theorem.
In our case, the measures provide independent descriptions of the system's susceptibility.
While the susceptibility measures peak at approximately the same noise levels for fixed sets of model parameters, we find that they respond differently to changes in the control parameters $\plink$ and $\prewire$. 
This is due to the fact that our definition for~$\chi$ does not strictly coincide with the usual statistical physics approach: 
The magnetic field is not directly present in the model, but we represent it as an additional stochastic process that can cause an agent to switch to a specific state.
For time development according to Glauber dynamics, this switching probability is a function of magnetization, external field and temperature.

By focusing on the susceptibility measures, we were able to identify three different adaptation pathways that allow the system to remain near the point of maximal susceptibility when external conditions change:
\begin{itemize}
    \item adjusting local degree by tuning the linking probability;
    \item changing individual compliance to the locally dominating opinion by tuning the non-linearity parameter;
    \item including or removing long-ranged correlation by tuning the rewiring probability.
\end{itemize}
Each pathway will have its own physical realisation on the level of individual agents.
For instance, while we chose~$b$ as identical for all agents, it is a parameter that \textit{acts} locally, so in principle, one could also allow individually different~$b$ values.

However, in order for the system to adapt autonomously, some knowledge about its macroscopic state is required on the microscopic level~\cite{romanczuk2023phase}.
Such feedback mechanisms have been demonstrated for neuronal systems, where adaptations based on spike-time dependent plasticity~\cite{meisel2009adaptiveselforganization} or homeostatic thresholds~\cite{menesse2022homeostaticcriticality} indeed allow interaction networks to self-organise to quasi-critical points.
In general, spin models are an important tool for understanding collective decision-making as an example of spontaneous symmetry breaking.
Here, we have provided further insights into the roles that network structure and non-linear social interactions play in this process.
However, fundamental questions remain open regarding the nature of feedback mechanisms that might facilitate the self-organised adaptation of collective systems in changing environmental contexts.

\section*{Acknowledgements}
We acknowledge support provided by the Deutsche Forschungsgemeinschaft (DFG, German Research Foundation) under Germany’s Excellence Strategy, EXC 2002/1 `Science of Intelligence’, Project 390523135 (Y.S. and P.R.) and National Research Foundation of Ukraine, Project 2023.03/0099 ``Criticality of complex systems: fundamental aspects and applications" (P.S. and Yu.H.) as well as a Richter Summer Fellowship and a Robin Berlin Fellowship granted through Yale University~(A.T.).

\FloatBarrier
\newpage
\bibliographystyle{ieeetr}
\bibliography{references.bib}


\end{document}

%% file: figures/schematic.tikz.tex
\begin{tikzpicture}[scale=0.7, node distance=2.5cm, every node/.style={circle, draw, minimum size=0.1cm}, 
    -]

    \node[draw, rectangle, minimum width=2cm, minimum height=1.4cm, align=center] (random) {Random\\Graph};
    \node[draw, rectangle, minimum width=2cm, minimum height=1.4cm, align=center] (complete) [left=of random, shift={(0,0.3)}]{Complete\\Graph};
    \node[draw, rectangle, minimum width=2cm, minimum height=1.4cm, align=center] (lattice) [right=of random, shift={(0,0.3)}] {2D Lattice\\Graph};

    \draw[->, thick] (complete) --  node[below,draw=none] {$\plink \to 0$} (random);
    \draw[->, thick] (lattice) -- node[below,draw=none] {$\prewire \to 1$} (random);

    \node[draw=none,fill=none] (randCenter) [below=of random, shift={(0,1.3)}] {};
    \node[draw=none,fill=none] (compCenter) [below=of complete, shift={(0,1.3)}] {};
    \node[draw=none,fill=none] (latCenter) [below=of lattice, shift={(0,1.3)}] {};

    \foreach \i in {1,...,9}
        \node[fill=black] (comp\i) at ($(compCenter) + (150+\i*40:1.3cm)$) {};
    \foreach \i in {1,...,9}
        \foreach \j in {\i,...,9}
            \draw (comp\i) -- (comp\j);
            
    \node[fill=black] (rand1) at ($(randCenter) + (-1.111, 0.492)$)  {};  
    \node[fill=black] (rand2) at ($(randCenter) + (0.640, -1.352)$)  {};  
    \node[fill=black] (rand3) at ($(randCenter) + (0.038, -0.620)$)  {};  
    \node[fill=black] (rand4) at ($(randCenter) + (1.048, 0.497)$)   {};   
    \node[fill=black] (rand5) at ($(randCenter) + (0.240, 0.180)$)   {};   
    \node[fill=black] (rand6) at ($(randCenter) + (-0.375, -1.402)$) {}; 
    \node[fill=black] (rand7) at ($(randCenter) + (1.088, -0.550)$)  {};  
    \node[fill=black] (rand8) at ($(randCenter) + (-0.304, 0.818)$)  {};  
    \node[fill=black] (rand9) at ($(randCenter) + (-0.881, -0.622)$) {};  

    \foreach \i/\j in {1/3, 1/5, 1/8, 2/6, 3/5, 3/6, 3/7, 3/9, 4/5, 4/8, 5/8, 6/9}
        \draw (rand\i) -- (rand\j);
        
    \foreach \x in {0,1,2}
        \foreach \y in {0,1,2}
            \node[fill=black] (lat\x\y) at ($(latCenter) + (\x-1, 1-\y)$) {};

    \foreach \y in {0,1,2} {
        \foreach \x in {0,1} {
            \pgfmathtruncatemacro{\xnext}{\x+1}
            \draw (lat\x\y) -- (lat\xnext\y);
        }
    }
    \foreach \x in {0,1,2} {
        \foreach \y in {0,1} {
            \pgfmathtruncatemacro{\ynext}{\y+1}
            \draw (lat\x\y) -- (lat\x\ynext);
        }
    }
    
\end{tikzpicture}